\documentclass{PoS}
\usepackage{bm}
\usepackage{url}
\usepackage{color}
\newcommand{\trento}{{\rm T\raisebox{-0.5ex}{R}ENTo}}

\title{Jet substructure modification in a QGP from a multi-scale description of jet evolution with JETSCAPE}

\ShortTitle{Jet substructure from multi-scale jet evolution with JETSCAPE}

\author{\speaker{Yasuki Tachibana} for the JETSCAPE Collaboration\\
        Department of Physics and Astronomy, Wayne State University\\
        Detroit, Michigan 48201, USA\\
        E-mail: \email{yasuki.tachibana@wayne.edu}}
\abstract{
The modification of jet substructure in relativistic heavy-ion collisions is studied using JETSCAPE, a publicly available 
software package 
containing a framework for 
Monte Carlo event generators. 
Multi-stage jet evolution in JETSCAPE 
provides an integrated description of jet quenching 
by combining multiple models, 
with each becoming active at a different stage 
of the parton shower evolution. 
Jet substructure modification 
due to different aspects of jet quenching 
is studied using 
jet shape and jet fragmentation observables. 
Various combinations of 
jet energy loss models 
are exploed, with medium background 
provided by 
$(2+1)$-D VISHNU 
with \trento+freestreaming initial conditions. 
Results reported here are from simulations performed within JETSCAPE framework. 
}

\FullConference{International Conference on Hard and Electromagnetic Probes of High-Energy Nuclear Collisions\\
		30 September - 5 October 2018\\
		Aix-Les-Bains, Savoie, France}

\begin{document}
\section{Introduction}
JETSCAPE (Jet Energy loss Tomography with a Statistically and Computationally Advanced Program Envelope) is a 
modular, flexible, publicly-released event generator framework 
for modeling heavy ion collisions~\cite{jsrepo,Kauder:2018cdt}.
In these proceedings 
we focus on 
the application of JETSCAPE to 
jet substructure observables, 
in particular jet shapes and jet fragmentation distributions, 
presenting results from numerical simulations. 
Utilizing multi-stage jet evolution incorporated in JETSCAPE,
we compare different combinations and settings of 
the jet energy loss modules 
and explore their sensitivity to different aspects of jet quenching. 
A complementary JETSCAPE study of 
$R_{\rm AA}$ and $v_2$ for jets and charged hadrons 
is presented in Ref.~\cite{Park:HP18}. 

\section{Multi-stage Jet Evolution in JETSCAPE}
Jet quenching is a multi-scale problem. 
JETSCAPE incorporates a multi-stage description of
jet evolution in matter~\cite{Cao:2017zih}, 
with a modular and flexible framework that enables 
different combinations of models for each stage. 
In this framework, 
the assignment and switching 
between stages is carried out 
for each parton, 
depending on its virtuality and energy during the propagation. 
In the current version of JETSCAPE, 
four different energy loss modules are available: 
MATTER, 
LBT, MARTINI, and AdS/CFT. 

MATTER~\cite{Majumder:2013re} 
is  a Monte Carlo (MC) event generator that simulates 
vacuum or medium-modified virtuality-ordered splittings of 
high energy partons with large virtuality $Q^2\gg \sqrt{\hat{q}E}$ 
(where $\hat{q}$ is the quenching transport coefficient, and 
$Q^2 > 1{\rm~GeV}^2$ in the vacuum). 
The Sudakov form factor is modified 
to include the effect of medium-induced radiation. 
This is then sampled to determine 
the presence or absence of each splitting 
and the virtuality of the parent parton 
at the time of its formation. 
The longitudinal momentum distribution 
of the daughter partons is determined 
using the combination of a vacuum and medium modified splitting functions. 
The effect of elastic scatterings with thermal partons in the medium 
is also taken into account. 
The scattered thermal partons (``recoils'') 
are transferred to the module 
that handles 
partons with small virtuality. 

The Linear Boltzmann Transport (LBT) model~\cite{Luo:2018pto} 
simulates 
in-medium high energy parton evolution 
at low virtuality scale, utilizing the 
linear Boltzmann equation. 
The elastic scattering contribution is calculated as 
$2\rightarrow2$ scatterings 
with thermal partons in the medium. 
The recoils from these scatterings evolve 
like all other low-virtuality partons. 
The inelastic process contribution is due to
medium-induced gluon radiation
and its spectrum 
is taken from the Higher Twist energy loss formalism. 

MARTINI~\cite{Schenke:2009gb} is 
an MC generator 
for in-medium transport of the hard partons with small virtuality 
based on the effective kinetic theory, 
including both elastic and inelastic processes. 
MARTINI employs 
the in-medium radiation rate of 
the AMY energy loss formalism~\cite{Arnold:2002ja, Arnold:2002zm}. 
The contribution of recoils 
is not included in this implementation of MARTINI in JETSCAPE. 
In this work, a momentum cut for hard partons 
$p_{\rm cut}=1~{\rm GeV}/c$ is applied. 

In the JETSCAPE AdS/CFT module~\cite{Casalderrey-Solana:2014bpa}, 
partons in jet showers are dragged 
assuming a non-perturbative strong interaction with the medium. 
The energy loss rate used in this module, 
that quantifies the amount of energy and momentum 
flowing from a jet parton to the medium hydrodynamic modes, 
is taken from the calculation 
for an $\mathcal{N}=4$ super Yang-Mills plasma 
in the limit of infinite coupling and large $N_c$
via the AdS/CFT correspondence \cite{Chesler:2014jva,Chesler:2015nqz}. 

Simulations of jet events 
in PbPb collisions at $2.76~A~{\rm TeV}$ 
are performed within JETSCAPE. 
The initial hard partons are generated 
by PYTHIA$+$\trento~\cite{Moreland:2014oya}. 
The hard partons then 
undergo medium-modified virtuality-ordered splitting generated 
by MATTER until their virtuality decreases to 
a specified value of separation scale, $Q_0=2~{\rm GeV}$. 
Partons are passed to another model for 
the small-virtuality stage, which is one of 
LBT, MARTINI, or AdS/CFT. 
For the vacuum shower evolution in pp collisions, 
MATTER without in-medium scattering is employed. 
Hadronization utilizes 
the Lund string model as implemented in PYTHIA. 
The event-averaged background medium profile in AA collisions
is calculated from $(2+1)$-D 
freestreaming pre-equilibrium evolution \cite{Liu:2015nwa}
and a viscous hydrodynamic calculation 
using VISHNU \cite{Shen:2014vra} 
with \trento\,initial conditions. 

\section{Results}

\begin{figure}
\begin{center}
\includegraphics[width=0.92\textwidth,bb=0 0 1024 400]{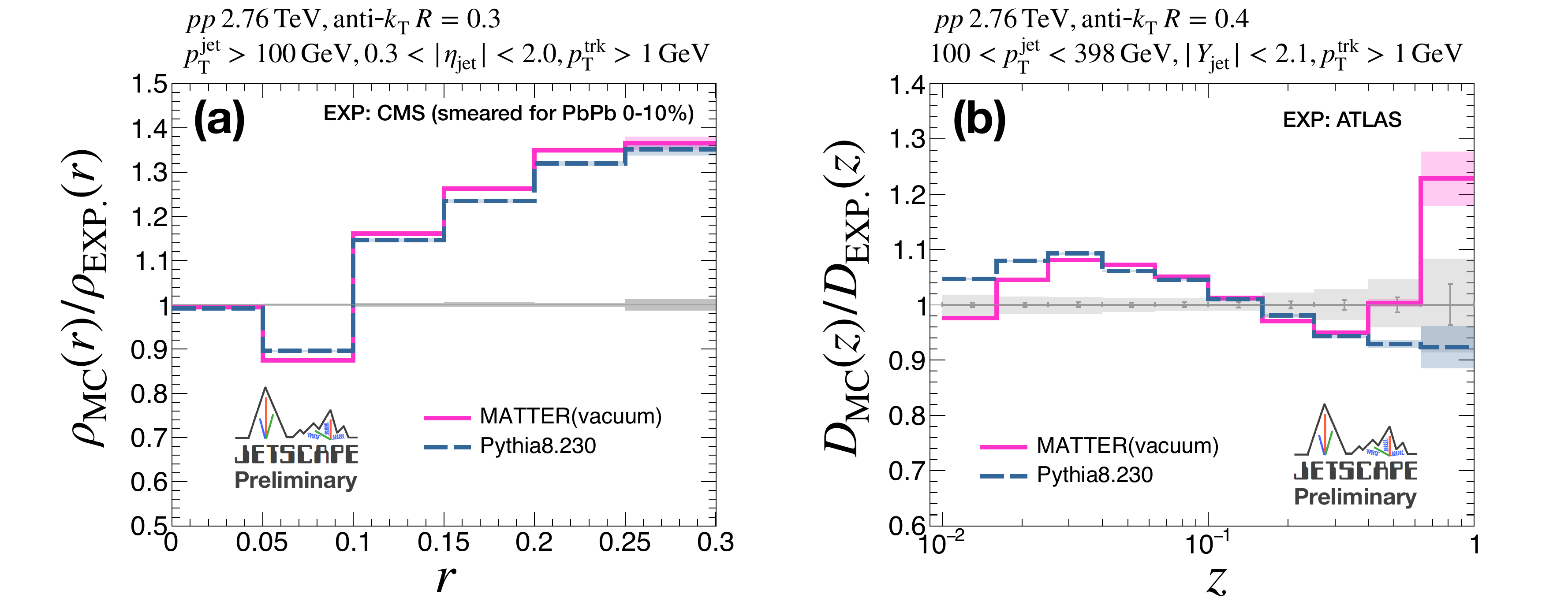}
\vspace{-8.5pt}
\caption{
Ratio of (a) jet shape and (b) jet fragmentation distribution in pp collisions, 
for JETSCAPE calculations 
and experimental data. 
The solid and dashed lines are for 
MATTER and PYTHIA. 
The experimental data are from {CMS}~\cite{Chatrchyan:2013kwa} and 
{ATLAS}~\cite{Aaboud:2017bzv}.
}
\label{fig1}
\end{center}
\end{figure}
Figure~\ref{fig1} shows the results for pp collisions. 
In the jet shape, 
the deficit in the second bin is correlated with 
deviations from the experimental data at large-$r$ 
due to the self-normalization in the definition of the observable. 
In the jet fragmentation distribution, 
a deviation at large-$z$ can be seen. 
The JETSCAPE-generated distribution for the jet shape 
is similar to that generated by
PYTHIA with default parameters, 
though deviations from 
PYTHIA can be seen for the jet fragmentation distribution. 
It should be noted that 
the fine tuning of parameters has not yet been carried out 
and is left for future work. 
\begin{figure}
\begin{center}
\includegraphics[width=.92\textwidth,bb=0 0 1024 400]{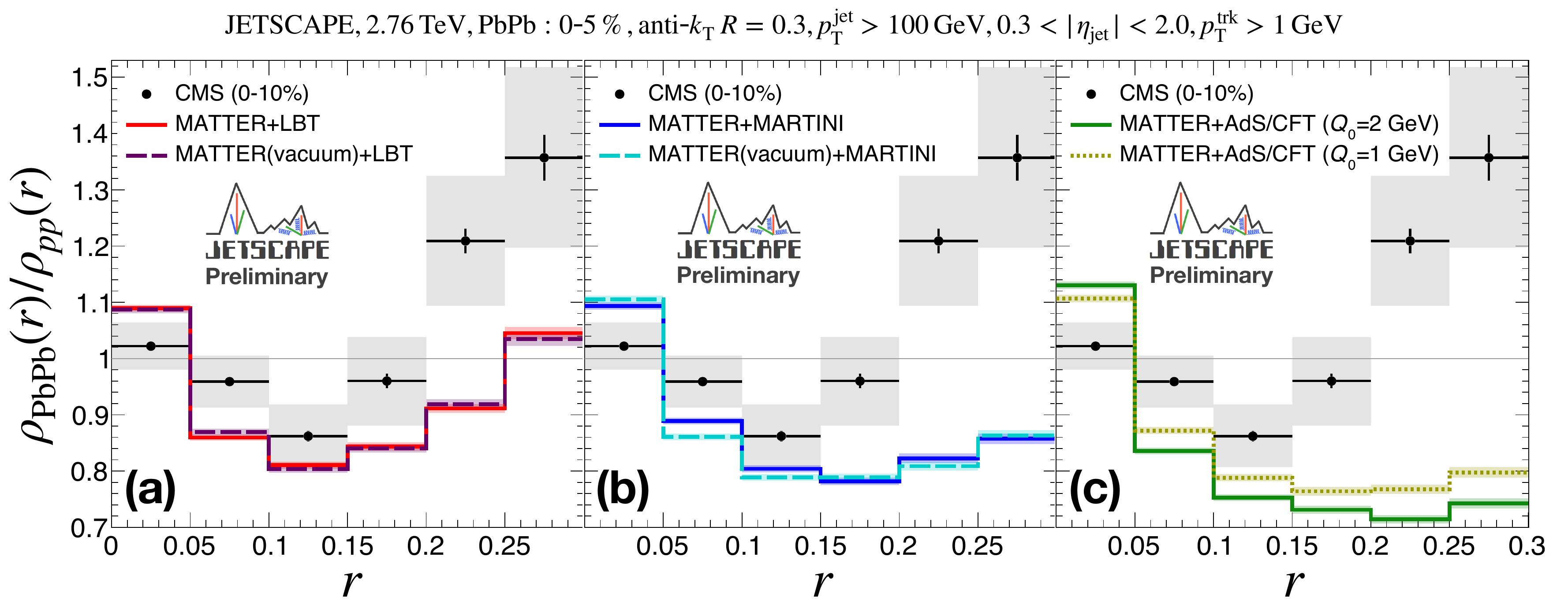}
\vspace{-5pt}
\caption{
PbPb/pp jet shape ratio 
from 
(a)~MATTER$+$LBT, 
(b)~MATTER$+$MARTINI, and
(c)~MATTER$+$AdS/CFT. 
The solid and dashed lines are 
with and without medium effects in MATTER. 
The dotted line is for MATTER$+$AdS/CFT with $Q_0 = 1~{\rm GeV}$. 
The experimental data are from {CMS}~\cite{Chatrchyan:2013kwa}.
}
\label{fig2}
\end{center}
\end{figure}
Figure~\ref{fig2} 
shows the ratio of the jet shape function for PbPb and pp collisions. 
All models show slightly more collimation 
in PbPb for $r<0.15$ than observed in the data. 
In MATTER$+$LBT, 
one can see the recoil effect 
as an enhancement around the edge of the jet cone. 
Given the self-normalized nature of this observable, 
the deviation of the models from data at larger $r$ 
is highly correlated with that at small $r$ 
and is not an independent check. 

\begin{figure}
\begin{center}
\includegraphics[width=.92\textwidth,bb=0 0 1024 400]{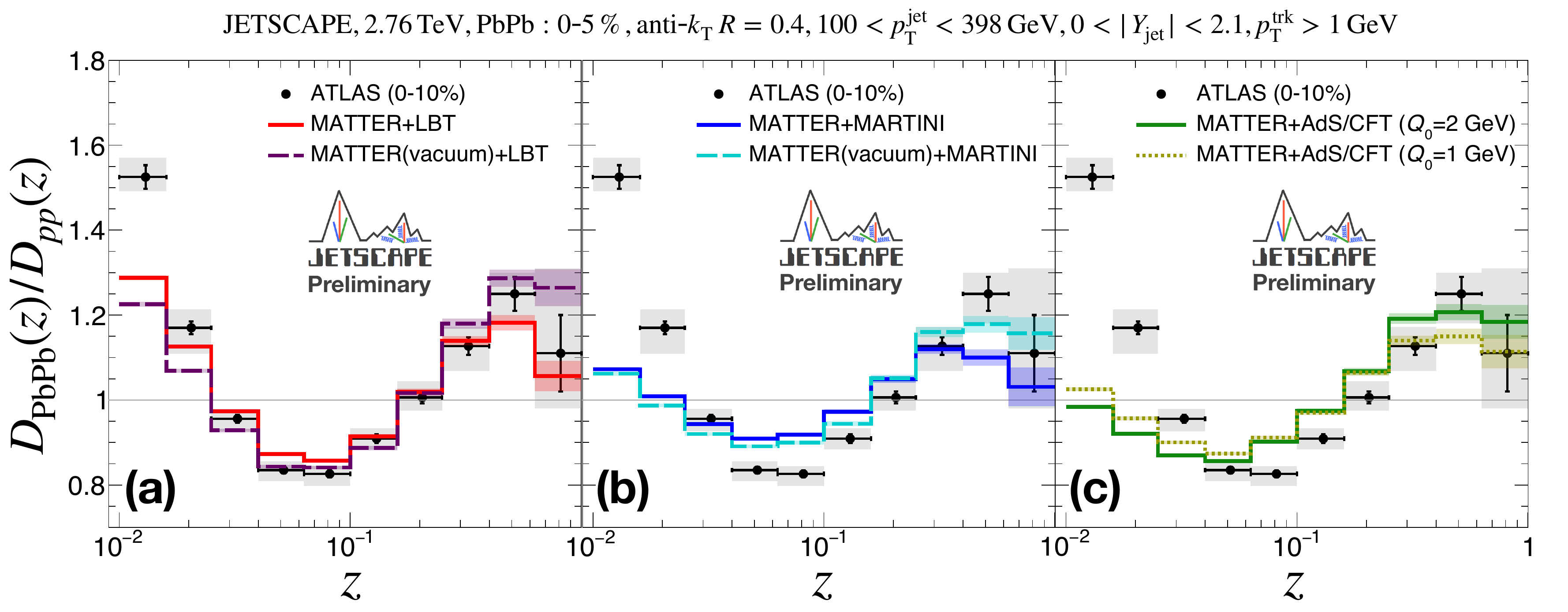}
\vspace{-5pt}
\caption{
PbPb/pp jet fragmentation distribution ratio 
from 
(a)~MATTER$+$LBT, 
(b)~MATTER$+$MARTINI, and
(c)~MATTER$+$AdS/CFT. 
The solid and dashed lines are 
with and without medium effects in MATTER. 
The dotted line is for MATTER$+$AdS/CFT with $Q_0 = 1~{\rm GeV}$. 
The experimental data are from {ATLAS}~\cite{Aaboud:2017bzv}.
}
\label{fig3}
\end{center}
\end{figure}
Figure~\ref{fig3} shows the
the ratio of the jet fragmentation distributions in PbPb and pp collisions. 
All combinations of model exhibit an enhancement 
at large-$z$. 
In contrast to jet shape function, 
the medium effect during the virtuality-ordered splitting 
can be seen clearly at large-$z$. 
In MATTER$+$LBT, 
one can see 
an enhancement at small-$z$ due to the recoils. 
This is more prominent when the medium effect in MATTER is on 
because the recoils in LBT inherit the medium effect in MATTER. 
Owing to the correlation between different $z$ 
of the jet fragmentation distribution, 
insufficient enhancements at small-$z$ 
due to the lack of recoils or medium response 
in MATTER$+$MARTINI and MATTER$+$AdS/CFT 
gives also deviations at intermediate-$z$. 

\section{Summary}
The modification of jet substructure due to jet quenching was studied with JETSCAPE, a newly developed event generator 
that employs a multi-stage description of the evolution of
the jet shower. 
The results for different combinations and settings 
of the jet energy loss modules were shown and compared. 
The collimation of jet shape at small-$r$ and 
the enhancement at large-$z$ 
are observed for all choices of model.
The medium effect during the virtuality-ordered splittings 
was found in the jet fragmentation distribution. 
Also we found that the recoils in the low-virtuality regime 
exhibited prominent effects; 
jet shape broadening and 
low-$z$ particle enhancements.

\vspace{5pt}
\begin{noindent}
\textbf{Acknowledgements:} 
These proceedings are supported in part by the National 
Science Foundation (NSF) within the framework of 
the JETSCAPE collaboration, under grant numbers ACI-1550300.
\end{noindent}


\bibliographystyle{JHEP}

          \bibliography{ref}


\end{document}